\documentstyle[preprint,aps,prd,eqsecnum,epsfig]{revtex}
\def\beqq{\begin{eqnarray*}}
\def\eeqq{\end{eqnarray*}}
\def\beq{\begin{equation}}
\def\eeq{\end{equation}}
\def\beqn{\begin{eqnarray}}
\def\eeqn{\end{eqnarray}}
\def\Z{Z \!\!\! Z}
\def\cZ{{\cal Z}}
\def\cC{{\cal C}}
\def\cL{{\cal L}}
\def\dd{{\rm d}}
\def\dD{{\rm D}}
\def\eq#1{(\ref{#1})}

\begin{document}
\draft
\preprint{ITEP-TH-29/00}
\title{Interaction of electric charges in (2+1)D magnetic dipole gas}
\author{M. N.Chernodub}
\address{Institute of Theoretical and Experimental Physics,\\
B.Cheremushkinskaya 25, Moscow, 117259, Russia}

\date{\today}
\maketitle
\begin{abstract}
The interaction of electrically charged particles in a dilute
gas of pointlike magnetic dipoles is studied. We show that the
interaction potential at small distances has a linear piece due to
overlap of the dipole clouds gathered near electric sources. At large
distances the potential becomes of the Coulomb type with
nonperturbatively renormalized charge of the test particle. The
physical applications of these results are discussed.
\end{abstract}


\narrowtext

\section{Introduction}

We study properties of electric charges in the plasma of the Abelian
magnetic monopoles in three dimensional Euclidean spacetime.
Physically, the magnetic dipoles may arise as the
monopole--antimonopole bound states in gauge theories like, for
example, the Georgi--Glashow model. This model possesses the
topologically stable classical solution called 't~Hooft--Polyakov
monopole~\cite{tHPo} which carry a unit of magnetic charge (the
instanton number in three dimensions). The 't~Hooft--Polyakov
monopole consists of a compact core and longrange gauge fields
associated with unbroken Abelian subgroup.

In the weak coupling regime the vacuum of the Georgi--Glashow model
is filled up with a dilute monopole--antimonopole plasma. Due to a
longrange nature of the Abelian fields the behavior of the plasma
is essentially the same as that of the Coulomb plasma of the Abelian
charges. Polyakov has shown~\cite{Polyakov} that in this plasma test
particles with opposite electric charges experience confining forces
at large separations due to formation of a stringlike object
between the charge and the anti-charge. The string has a finite
thickness of the order of the plasma correlation length ({\it i.e.},
the inverse Debye mass) and a finite energy per unit of the string
length ("string tension"). Thus the potential between the test
particles is linear at large distances~\cite{Polyakov}.

Small distance behavior of the inter-particle potential in the
monopole gas is also interesting.  According to
Ref.~\cite{ChGuPoZa:GG} at distances $R$ much smaller than the
correlation length of the plasma the potential contains a
nonperturbative piece proportional to $R^\alpha$, $\alpha \approx
0.6$ in addition to a perturbative contribution due to the one photon
exchange. The former is a consequence of the string formation at large
distances. A nontrivial short distance potential may have many
physically interesting consequences, see, {\it e.g.}
Ref.~\cite{VIZandCo,GuPoZa98-99} for a discussion in context of QCD
and other theories.

At a sufficiently high temperature the Coulomb plasma was shown to
undergo the Be\-re\-zin\-sky--Kos\-ter\-litz--Thou\-less phase
transition~\cite{B,KT}. In the  high temperature phase the charges
form neutral bound states~\cite{KT,FrSp80,AgasianZarembo} which obey
nonzero dipole moments.  Thus at the phase transition
the Coulomb plasma of charges transforms into a gas of the dipoles in
which the Debye screening is absent~\cite{no-screening,FrSp80}. The
field of the magnetic dipole is too weak at large distances to be
able to induce a nonzero string tension between the electric
charges. As a consequence of this fact, at large separations between
the electric charges the string structure is destroyed in the high
temperature phase and the potential becomes of the Coulomb type.  A
detailed study of the temperature phase transition in the three
dimensional Georgi--Glashow model can be found in
Ref.~\cite{AgasianZarembo}. The physics of dipoles in gauge theories
has also been discussed in Ref.~\cite{Antonov}

Despite of the absence of the longrange confinement the physics of
the electric charges in the dipole gas is still interesting. In this
paper we show that the short range potential between the charges has
a linearly rising piece due to interaction of the electric charges
with the magnetic dipoles.  For simplicity we consider a dilute gas
of the pointlike dipoles characterized by a (fluctuating in general
case) dipole moment $\vec \mu$ and the spacetime position $\vec x$.
The temperature effects are not studied in the present publication.
The structure of the paper is as follows. The path integral
formulation of the dilute dipole gas is presented in Sec.\
\ref{section:path}.  The interaction of the magnetic monopole gas
with the electric charges is discussed in Sec.\ \ref{section:int} and
interaction energy of the static charge--anticharge pair is studied
in Sec.\ \ref{section:potential}. The generalization of the present
approach to the fluctuating dipole moments is given in Sec.\
\ref{fluct:moments}.  Our conclusions are summarized in the last
section.

\section{Path integral formulation}
\label{section:path}

The magnetic dipole is a magnetically neutral localized pair of monopole and
anti-monopole separated by the distance $\vec r$. The magnetic moment of the
dipole is $\vec\mu = g_m \cdot \vec r$ where $g_m$ is the magnetic charge of the
constituent monopole and $\vec r$ is the distance between the constituents.
If the typical distance between the
dipoles is much larger than the dipole size
$r$ then the dipoles may be treated as pointlike particles. This condition
can be written as follows:
\beq
\rho^{\frac{1}{3}} \, r \ll 1\,,
\label{diluteness}
\eeq
where $\rho$ is the mean density of the dipole gas. In this case
the dipole is characterized solely by the position $\vec x$ and the dipole
moment $\vec \mu$ while the "dipole size" $r$ becomes an internal
characteristic of the dipole. The action of two interacting pointlike
dipoles with magnetic moments $\vec\mu_a$ and $\vec\mu_b$ located at
positions $\vec x_a$ and $\vec x_b$, respectively, is given by the formula:
$$
V(\vec\mu_a,\vec\mu_b;\vec x_a,\vec x_b) =
(\vec \mu_a \cdot \vec \partial)\, (\vec \mu_b \cdot \vec \partial)
D_{(3D)}(\vec x_a,\vec x_b)\,,
$$
where $D_{(3D)}(x) = {(4 \pi |x|)}^{-1}$ is the propagator for a scalar
massless particle in three dimensions.

The statistical sum of the dilute dipole gas can be written as follows:
\widetext
\beq
\label{GasPF}
\cZ = \sum\limits^\infty_{N=0} \frac{\zeta^N}{N!}
\int \dd^3 x_1 \int\dd^3 \mu_1 \cdots
\int \dd^3 x_N \int\dd^3 \mu_N
\exp\Biggl\{ - \frac{1}{2} \sum^N_{\stackrel{a,b=1}{a \neq b}}
V(\vec\mu_a,\vec\mu_b;\vec x_a,\vec x_b)\Biggr\}\,,
\eeq
where $\zeta$ is the fugacity parameter \footnote{Note that in $2+1$
theory the dimensionalities are:  $[g^2_m] = {\rm{mass}}^{-1}$,
$[\mu] = {\rm{mass}}^{-\frac{3}{2}}$ and
$[\zeta] = {\rm{mass}}^3$.}. As in the case of the monopole
gas~\cite{Polyakov} the dipole fugacity is a nonperturbative quantity
since $\zeta \sim e^{-S_0}$, where $S_0 \sim g^{-2}_e$ is
the action of a single dipole  and $g_e = 2 \pi \slash g_m$ is the
fundamental electric charge in the theory.
\narrowtext

The integration over the dipole moment $\int \dd^3 \mu$ in
eq.\eq{GasPF} is given by the integration over direction $\vec n$ of
the dipole moment $\vec \mu_a = \mu_a \cdot \vec n_a$ (no sum is
implemented) and over its absolute value, $\mu_a$, weighted with a
distribution function. For the sake of simplicity we fix the absolute
value of all magnetic  moments $\mu_a = \mu$. The case of the
fluctuating $\mu$ is considered in Sec.\ \ref{fluct:moments}.

Introducing the scalar field $\chi$ we rewrite expression \eq{GasPF}
as follows:
\beqn
\exp\Bigl\{ - \sum^N_{\stackrel{a,b=1}{a \neq b}}
V_{ab}\Bigr\} & \propto &
\int\limits^{+\infty}_{-\infty} \dD \chi
\exp\Bigl\{- \frac{1}{2} \int \dd^3 x \,
{(\vec \partial \chi)}^2 \nonumber\\
& & + i \mu \sum\limits^N_{a=1}
(\vec n_a \cdot \vec \partial) \chi(x_a)\Bigr\}\,,
\label{step1}
\eeqn
where $(\vec a,\vec b)$ is the scalar product of the
vectors $\vec a$ and $\vec b$.
Substituting eq.\eq{step1} into eq.\eq{GasPF} and
summing over all dipole positions, we get the following representation of
the dipole partition function:
\widetext
\beqn
\cZ & \propto & \int \dD \chi \, \exp\Bigl\{ - \frac{1}{2} \int \dd^3 x \,
{(\vec \partial \chi)}^2 \Bigr\} \cdot
\sum\limits^\infty_{N=0} \frac{\zeta^N}{N!}
{\Biggl(
\int \dd^3 x \int\limits_{\vec n^2=1} \!\!\!\! \dd^3 n \,
\exp\{i \mu (\vec n \cdot \vec \partial) \chi(x)\}
\Biggr)}^N \nonumber\\
& = & \int \dD \chi \, \exp\Bigl\{ - \int \dd^3 x \,
\Bigl[\frac{1}{2} {(\vec \partial \chi)}^2 - \zeta \!\!
\int\limits_{\vec n^2=1} \!\!\dd^3 n \, \exp\{i \mu
(\vec n \cdot \vec \partial) \chi(x)\}\Bigr]
\Bigr\}\,.
\label{GasPF2}
\eeqn
\narrowtext

After integration over the vector $\vec n$ we obtain (up to an
inessential multiplicative factor):
\beq
\cZ = \int \dD \chi \, \exp\Bigl\{ - \int \dd^3 x \,
\cL[|\vec \partial \chi|] \Bigr\}\,,
\label{PF:basic}
\eeq
where $|\vec \partial \chi| = \sqrt{{\vec \partial \chi}^2}$, and
\beq
\cL(f) = \frac{1}{2} f^2 - 4 \pi \zeta \, v(\mu f)\,,\quad
v(f) = \frac{\sin f}{f}\,.
\label{V}
\eeq

We treat the model (\ref{PF:basic},\ref{V}) as an effective theory.
Shifting the field $f = |\vec \partial \chi|$,
$f \to f \slash \mu = \bar f$
and rescaling the space coordinates,
$x \to {(4 \pi \zeta)}^{-{1 \over 3}} x$
we immediately determine that the
dynamics of the model is controlled by
the dimensionless constant
\beq
\lambda=4 \pi \zeta \mu^2\,, \label{lambda}
\eeq
and the typical field fluctuation is of the order of ${\bar f}^2 \sim
\lambda$. Thus at small values of the parameter $\lambda$
lagrangian \eq{V}
can be expanded into series of $f^2$ and the interaction terms
$f^{2n}$ with $n  \geq 2$ can be treated as small perturbations.

The vacuum expectation value, $\rho = <\rho(x)>$,
of the dipole density,
$$
\rho(x) = \sum\limits_i \delta^{(3)} (x - x_i)\,,
$$
can be rewritten in representation~\eq{GasPF2}:
\beq
\rho =  4 \pi \zeta \, \Bigl<\frac{\sin(|\vec \partial
\chi(x)|)}{|\vec \partial \chi(x)|}\Bigr> = 4 \pi \zeta
\Bigl(1 + O(\lambda)\Bigr)\,,
\label{dens:quant}
\eeq
where the last equality is written for small couplings $\lambda$.
In this regime the coupling $\lambda$ is proportional to the
density of the dipoles, $\lambda = \rho \,  \mu^2 + O(\lambda^2)$,
and therefore the condition $\lambda \ll 1$ can be interpreted as a
requirement for the density of the dipole moments to be small.

\section{Electric charges in magnetic dipole gas}
\label{section:int}

An infinitely heavy test particle carrying the electric charge $q g_e$
in introduced in the vacuum of the theory with the help of the
Wilson loop operator, $W_q (\cC)$. Here contour $\cC$ is the trajectory
of the particle. Let us first consider the contribution of the Abelian
monopoles to the Wilson loop:
\beq
W_q(\cC) = \exp\Bigl\{i \int \dd^3 x \,
\rho_{\rm{mon}}(x) \eta^\cC(x)\Bigr\}\,,
\label{WLq}
\eeq
where $\rho_{\rm{mon}}(x) = g_m \, \sum_a m_a \delta(x - x_a)$,
$m_a \in \Z$ is magnetic charge density. The function $\eta_\cC$ is
defined as follows:
$$
\eta^\cC(x) = \pi \int\limits_{\Sigma_\cC} \dd^2 \sigma_{ij}(y) \,
\varepsilon_{ijk} \, \partial_k D(x-y)\,,
$$
where the integration is taken over an arbitrary surface $\Sigma_\cC$
spanned on the contour $\cC$. Note that the value of the Wilson loop does not
depend on the shape of the surface $\Sigma_\cC$.

Now suppose that all monopoles and anti-monopoles appear always as dipole
pairs of the small size $r$. In the pointlike dipole approximation
eq.\eq{WLq} can be rewritten as follows\footnote{The function $\eta_\cC$ has
the jump $\eta_\cC \to \eta_\cC + 2 \pi$ at the position of the surface
$\Sigma_\cC$. This jump is inessential for our considerations due to
$2\pi$--periodicity of the exponential function of an imaginary argument.}:
\beqn
W_q(\cC) & = & \exp\Bigl\{i \, r \, q \, \int \dd^3 x \,
(\vec\rho^{(\mu)}(x), \vec\partial) \eta^\cC(x)\Bigr\}\,, \nonumber\\
       & = & \exp\Bigl\{i \, r \, q \, \int \dd^3 x \,
(\vec\rho^{(\mu)}(x), \vec\gamma^\cC(x))\Bigr\}\,,
\label{WL}
\eeqn
where $\vec\gamma^\cC = \vec \partial \eta^\cC$, and
$\vec\rho^{(\mu)}(x) = \sum\nolimits_i \vec \mu_i \delta^{(3)} (x -
x_i)$ is the density of the dipole moments. Since the charge of the
Wilson loop $q$ always
enters eq.\eq{WL} in the combination $q r$ we consider below
the Wilson loop for the unit charge, $W(\cC) \equiv W_1(\cC)$, and then
make the rescaling $r \to q r$ in the final result to describe the potential
for $q \neq 1$.

To study the static potential between charges we consider the flat Wilson
loop operator defined in the plane $(x_1,x_2)$, see Fig.\ \ref{WLplane}.
The operator inserts the static charge and anti-charge pair separated by the
distance  $R$. The particle trajectory consists of the two straight lines,
$\cC = L_1 + L_2$ and the surface $\Sigma_\cC$ is located in the plane
$(x_1,x_2)$. The corresponding function $\eta^\cC$ is:
$$
\eta^\cC(x) = {\rm {arctan}} \Bigl(\frac{x_1 + R \slash 2}{x_3}\Bigr)
- {\rm {arctan}} \Bigl(\frac{x_1 - R \slash 2}{x_3} \Bigr)\,,
$$
and the function $\vec \gamma^\cC$ is defined as follows:
\beqn
\vec \gamma^\cC (x) & = & \pi \int\limits_{\Sigma_\cC}
\dd^2 \sigma_{ij}(y) \varepsilon_{ijk}\, \frac{\partial}{\partial x_k}
\vec \partial_x D_0(x-y)
\label{gamma:general}\\
& = & \vec \gamma^\cC_{\rm{sing}}(x)
+ \vec \gamma^\cC_{\rm{reg}}(x)\,;
\nonumber\\
\gamma^\cC_{{\rm{sing}},i}(x) & = & \pi \,
\delta_{i3} \, \delta(x_3)\, \Theta(2 x_1 + R)\,
\Theta(2 x_1 - R) \,; \label{gamma:singular}\\
\vec \gamma^\cC_{\rm{reg}}(x) & = & 4 R \,
\Bigr[ 8 \, x_1 \, x_3, \,\,0
\,\,,\,\, R^2 + 4 (x^2_3-x^2_1)\Bigl] \nonumber\\
& & \times {\Bigl[ {\Bigl(R^2 + 4 (x^2_3+x^2_1)\Bigr)}^2
- 16 R^2 x^2_1 \Bigr]}^{-1}\,,\nonumber
\eeqn
where $\Theta(x) = 1$ if $x>0$ and $\Theta(x) = 0$ otherwise.

Substituting eq.\eq{WL} into eq.\eq{GasPF} and
performing the transformations presented in
Sec.\ \ref{section:path} we represent the
quantum average for the Wilson loop in the dilute gas as follows:
\beqq
{<W(\cC)>}_{\rm{gas}} & = & \frac{1}{\cZ} \int \dD \chi \,
\exp\Bigl\{ - \int \dd^3 x \, \Bigl[
\frac{1}{2} (\vec \partial \chi)^2 \\
& & + 4 \pi \zeta\, \Bigl( 1 -
v(| \mu \,\vec \partial \chi + r \,\vec \gamma^\cC|)
\Bigr)\Bigr]\Bigr\}\,,
\eeqq
where the expression in the square brackets is normalized to zero
at $\chi=0$ and the function $v$ is defined in eq.\eq{V}.

Physically, the magnetic dipole gas affects the potential between
oppositely charged electric particles because of overlapping of the
dipole clouds gathered near the electric sources. Indeed, from the
point of view of the three dimensional physics the static test charge
trajectories $\cC= L_1 + L_2$, Fig.\ \ref{WLplane}, may be
considered as an electric current running along the contour $\cC$.
This current induces a magnetic field which circumvents each of the
paths $L_1$ and $L_2$.  Since the magnetic dipole lowers its energy
in the magnetic field, the density of the magnetic dipoles should
increase towards the position of the test electric charges. The
interaction energy of the dipole clouds gathered near the oppositely
charged particles depends on the inter-particle separation. Below we
show that at small charge separations this energy is a linear
function of the distance between the charges.

\section{Potential of static charge--anticharge pair}
\label{section:potential}

The static interaction potential of particles with electric charges
$\pm q g_e$ is a sum of the perturbative
contribution from the one photon exchange and the contribution from the
dipole gas, respectively:
\beqn
V_q(R) & = & - \frac{q^2 g^2_e}{2}\, D_{(2D)}(R)
+ E^{\rm{gas}}_q(R) \,,\nonumber\\
E^{\rm{gas}}_q(R) & = & - \lim\limits_{T \to \infty}
\frac{1}{T} <W_q(\cC_{R\times T})>\,,
\label{full:potential}
\eeqn
where $D_{(2D)}(R) = - {(2 \pi)}^{-1} \, \log(mR)$ is the
two dimensional propagator for a scalar massless particle, $m$ is
the massive parameter and $\cC_{R\times T}$ stands for the
rectangular $R\times T$ trajectory of the test particle. In the limit
$T \to \infty$ the contour $\cC_{R\times T}$ transforms into the
contour $\cC = L_1 + L_2$, Fig.\ \ref{WLplane}.

We evaluate the dipole gas contribution $E^{\rm{gas}}_q$ classically
minimizing the functional (we put $q=1$),
$$
S(\chi) = \int \dd^3 x \, \Bigl[ \frac{1}{2} (\vec \partial \chi)^2
+ 4 \pi \zeta\, \Bigl(1 - \frac{\sin
(|\mu \, \vec \partial \chi + r \, \vec \gamma^\cC |)}{
| \mu \, \vec \partial \chi + r \, \vec \gamma^\cC|} \Bigr)\Bigr]\,,
$$
with respect to the field $\chi$ in the limit $\lambda \ll 1$.
To this end we note that the singular part \eq{gamma:singular} of the
source \eq{gamma:general} does not contribute to the functional
$S(\chi)$. This statement can be checked directly
either by regularizing the singular piece of the source $\vec \gamma$ at
a finite $R$ or by considering the limit $R\to \infty$ in which only the
singular part $\gamma^\cC_{{\rm{sing}}}$ is nonzero at finite $x_1$
and $x_3$. Applying rescaling arguments similar to those mentioned after
eq.\eq{V} we conclude in the limit $\lambda \ll 1$ the classical
solution is suppressed as
$|\vec \partial \chi_{\rm{cl}}| = O(\lambda)$ and
therefore in the leading order in $\lambda$ the classical energy in the
leading order is given by the formula:
\beqn
E^{\rm{gas}}_{\rm{cl}}(R)
& = & 4 \pi \zeta \int \!\dd x_1 \int \!\dd x_3 \,
\Bigl(1 - v(r \, \Gamma^\cC(x,R)) \Bigr)\Bigr]\,,
\label{en}\\
\Gamma^\cC(x,R) & = & 4R \,
{\Bigl[{\Bigl(R^2 + 4 (x^2_3+x^2_1)
\Bigr)}^2 - 16 R^2 x^2_1\Bigr]}^{-\frac{1}{2}}\!.
\label{Gamma}
\eeqn
Unfortunately the explicit analytical integration is impossible due to the
complicated
form of the function $\Gamma$, eq.\eq{Gamma}. However, if the
distance $R$ between the test particles is much smaller than the "dipole
size" $r$ then the energy $E_{\rm{cl}}(R)$ can be expanded in series of
$R \slash r$:
\beq
E^{\rm{gas}}_{\rm{cl}}(R) = \pi^3 \zeta r R \, \Bigr(1 +
O\Bigl(\frac{R}{r}\Bigr)\Bigr)\,,\quad R \ll r\,.
\label{Ecl:short}
\eeq
Thus, the energy of the charge--anticharge pair is the linearly
rising function of the inter-particle distance

Note that in realistic theories the magnetic dipoles are realized
as mo\-no\-po\-le--anti\-mo\-no\-po\-le bound states with a nonzero 
{\it physical} size $r$. If the distance between test particles $R$ 
is much smaller than the size of the such dipole, $R \ll r$, then the 
test particles become affected by the monopole constituents and the
pointlike approximation to the dipole is no more valid. However as
we will see in the next Section the linear potential extends over
$R\sim$~(a~few)~$r$ and therefore we believe that our results derived
in the pointlike dipole approximation may also be applicable to
real systems.

According to eq.(\ref{Ecl:short}) the coefficient $\sigma_q$ in front
of the linear term in the interaction potential of the particles
carrying the electric charges $\pm q g_e$ is:
\beq
\sigma_q = q \, \sigma_1\,,\quad \sigma_1 = \pi^3 \zeta r =
\frac{\pi^2}{4} \, \rho \, r\,,
\label{sigma:q}
\eeq
where $\rho$ is the dipole density, eq.\eq{dens:quant}. This formula is
given in the leading order of $\lambda$. To derive eq.\eq{sigma:q}
we have used rescaling $r \to q r$ according to the discussion
after eq.\eq{WL}.

The coefficient $\sigma_q$ is proportional to the electric charge of
the test particle. Thus, each elementary electric "flux" coming
from the charge is not interacting with the other fluxes and, as a
result, the total coefficient, $\sigma_q$, is a sum of the elementary
coefficients $\sigma_1$ of $q$ individual fluxes. This is due to the
fact that leading contribution to the formation of the dipole clouds
is given by the external magnetic field induced by the test
particles. The dipole dynamics would affect the dipole clouds in next
to the leading order terms in expansion over the $\lambda$ parameter.
Since the density of the dipole clouds is a function of the test
particle charge, then the dipole density--density interactions in
higher order in $\lambda$ would lead to $q^n$, $n>1$ corrections to
the linearity coefficient $\sigma_q$. Thus it is natural to suggest
that in general case the proportionality of the linear term to
the electric charge of the test particles is lost even on the
classical level. Note that the proportionality of the linear term
coefficient $\sigma$ to the number of elementary fluxes appears
in the so called Bogomol'ny limit~\cite{Bogomolny} of the Abelian
Higgs model for the Abrikosov--Nielsen--Olesen strings~\cite{ANO}.
However, in the latter case the proportionality of the string tension
to the number of fluxes is exact on the classical level.

At large distances, $R \gg r$, the dipole gas contribution to
the interaction energy (\ref{en},\ref{Gamma}) between the
oppositely charged electric particles grows logarithmically:
\beq
E^{\rm{gas}}_{\rm{cl}}(R) = \frac{8 \pi^2}{3}
\zeta r^2 \log\Bigl(\frac{R}{r}\Bigr)
\, \Bigr[1 + O\Bigl(\frac{r}{R}\Bigr)\Bigr]\,,\quad R \gg r\,,
\label{Ecl:long}
\eeq
in accordance with absence of the charge screening in the dipole
gases~\cite{no-screening,FrSp80}. Shifting the dipole size $r \to q
\, r$ in eq.\eq{Ecl:long} we conclude that the dependence of the
nonperturbative potential \eq{Ecl:long} on charge of the test particles,
$q$, and on the distance between them, $R$, is essentially the same as
for the perturbative one photon exchange, eq.\eq{full:potential}.
Thus the dipole gas nonperturbatively renormalizes the coupling
constant $g_e$ at large distances and the full potential
\eq{full:potential} has the following behavior:
\beqq
V_q(R) & = &
\frac{\epsilon q^2 g^2_e}{4 \pi}\, log(R) + const.\,,\\
\epsilon & = & 1 + \frac{1}{3}\, \lambda + O(\lambda^2)\,,\quad R
\gg r\,,
\eeqq
where $\epsilon$ is the dielectric constant.  Note
that both the appearance of the linear potential at small distances
and the renormalization of the electric charge at large distances, $
g^2_e \to \epsilon g^2_e $, are essentially nonperturbative effects
since $\lambda \sim e^{- const. \slash g^2_e}$.

\section{Fluctuating dipole moments}
\label{fluct:moments}

In previous Sections we considered the properties of the "rigid" dipoles
which are characterized by a {\it fixed} absolute value of the
magnetic dipole
moment $\mu$. In a realistic case the dipole is realized as a
monopole--antimonopole bound state with a fluctuating absolute value of
the dipole moment. In this Section we generalize our approach
to the case of the fluctuating dipole moments $\mu = g_m \, r$.  It is
convenient to describe the fluctuations of the dipole moments as the
fluctuations of the dipole size $r$ at a fixed magnetic charge $g_m$ of
the monopole constituents. In a general case the distance $r$
can be characterized by a distribution function $F(r)$ normalized to
unity, $\int^{+\infty}_0 \, \dd \, r \, F(r) = 1$.
We assume that the dipole gas is sufficiently
dilute ({\it cf.} eq.\eq{diluteness}):
${<\rho_D>}^{\frac{1}{3}} \int \dd r \, r\, F(r) \ll 1$.

The statistical sum of the gas of pointlike dipoles
with the fluctuating moments is:
\widetext
\beq
\cZ =
\sum\limits^\infty_{N=0} \frac{\zeta^N}{N!}
\prod\limits^N_{i=1} \Bigl(
\int\limits^{+\infty}_0 \dd r_i \, F(r_i) \int \dd^3 x_i
\!\! \int\limits_{{\vec n}^2_i=1} \!\! \dd^3 n_i \Bigr) \,
\exp\Biggl\{ - \frac{1}{2} \sum^N_{\stackrel{a,b=1}{a \neq b}}
V(g_m\,r_a\,{\vec n}_a, g_m\,r_b\, {\vec n}_b;
{\vec x}_a, {\vec x}_b) \Biggr\}\,.
\label{GasPF:fluct}
\eeq
\narrowtext
In the case of a fixed dipole moment, $F(r) = \delta(r - d)$, the
above  sum is reduced to the partition function \eq{GasPF} with $\mu
= g_m \, d$.

Analogously to Section~\ref{section:path} we introduce the field $\chi$,
perform summation over the dipole positions $x_a$, and integrate out the
dipole directions $\vec n$. The partition function \eq{GasPF:fluct}
is then represented as follows:
$$
\cZ = \int \dD \chi \, \exp\Bigl\{ - \int \dd^3 x \,
\Bigl[ \frac{1}{2} {(\vec\partial \chi)}^2 - 4 \pi \zeta \,
v_{\rm{fl}}(g_m |\vec \partial \chi|) \Bigr] \Bigr\}\,,
$$
where
\beq
v_{\rm{fl}}(f) = \int\limits^{+\infty}_0 \dd \, r \, F(r)\,
\frac{\sin (r\, f)}{r \, f}\,.
\label{Vfl}
\eeq
Note that the action for the field $\chi$ has an additional integration
over the dipole size parameter $r$ in comparison with
eqs.(\ref{PF:basic},\ref{V}).

As a simple example let us consider the following distribution function,
shown in Fig.\ \ref{fluct:fig}(a):
\beq
F(r) = \frac{r}{d^2} \exp\Bigl\{ - \frac{r}{d}\Bigr\}\,,
\label{distr}
\eeq
which has a maximum at $r=d$, and the mean value of the dipole size,
$r$, is $\bar r = 2 d$.

Function \eq{Vfl} for distribution \eq{distr} has the following form,
$v_{\rm{fl}}(f) = \frac{1}{1 + d^2 \, f^2}$,
and the expansion parameter $\lambda_{\rm{fl}}$ is now given by
eq.\eq{lambda} with $\mu = {\bar r} \, g_m \equiv 2 d \, g_m$:
$$
\lambda_{\rm{fl}} = 16 \pi \zeta d^2 g^2_m\,.
$$
In the leading order in $\lambda_{\rm{fl}}$ the
classical inter-particle energy can be calculated similarly to the
previous Section:
\beqn
E^{\rm{gas,fl}}_{\rm{cl}}(R) = 4 \pi
\zeta \, \int \dd^2 x \, w(x,R) = 2
\pi^3 \zeta d^2 \, G\Bigl(\frac{R}{d}\Bigr)\,, \label{E}\\
w(x,R) = \frac{d^2 \Gamma^2_\cC(x_1,x_3,R)}{1 + d^2 \,
\Gamma^2_\cC(x_1,x_3,R)}\,,
\label{energy:density}
\eeqn
where $\Gamma^\cC$ is defined in eq.\eq{Gamma}.

The leading behavior of the energy $E^{\rm{gas,fl}}_{\rm{cl}}$ at
large and small distances is given by the following formulae,
respectively:
\beqn
E^{\rm{gas,fl}}_{\rm{cl},q}(R)  = 2 \pi^3 \zeta
d \, q \, R= \frac{\pi^2}{2} \, \rho \, d\,q\,R\,,\quad R \ll r\,,
\label{Efl:short}\\
E^{\rm{gas,fl}}_{\rm{cl},q}(R) = 16 \pi^2 \zeta d^2 q^2 \,
\log \Bigl(\frac{R}{r}\Bigr)\,, \quad R \gg r\,, \nonumber
\eeqn
where the dipole density in the leading order in $\lambda_{\rm{fl}}$
is given in eq.\eq{dens:quant}. As in the case of
the ordinary dipoles the presence of the dipole
gas leads to the appearance of the linear term in the inter-particle
potential. The coefficient in front of the linear term is:
$$
\sigma^{\rm{fl}}_q = q\,
\sigma^{\rm{fl}}_1\,,\quad
\sigma^{\rm{fl}}_1 = 2 \pi^3 \zeta d\,.
$$
At large distances the renormalization of the electric charge
takes place:
\beq
g^2_e \to \epsilon_{\rm{fl}} g^2_e \, \,, \quad
\epsilon_{\rm{fl}} = 1 + \frac{1}{2} \lambda_{\rm{fl}} +
O(\lambda_{\rm{fl}}^2)\,.
\eeq

The profile and contour plots for the (normalized) energy density $w$,
eq.\eq{energy:density} of the dipole cloud are shown in
Fig.\ \ref{fig:density} for various charge separations $R$. Although at
small distances the profiles show some similarity with a stringlike
object the width of such a "string" grows as $R_{str} \approx
\sqrt{d\,R}$ for $R \ll d$. Indeed, the effective width $R_{str}$
of the string profile at the center of the string can be defined as
follows:
\beq
R^2_{str} = \frac{\int\limits^{\infty}_{-\infty} x^2_3 \, w(0,x_3,R)
\dd x_3}{\int\limits^{\infty}_{-\infty} \, w(0,x_3,R) \dd x_3} =
\frac{R \sqrt{R^2+16 d^2}}{4}\,.
\eeq
Since the energy of the "string" is proportional to the volume it
occupies the "string" could merely be responsible for $R^2_{str}(R)
\cdot R \sim R^2$ (and not for $\sim R$) term in the potential. Thus
the linear term in the potential has nothing to do with the observed
string formation.

It is instructive to study the behavior of the (normalized) energy of
the test particle pair $G$, eq.\eq{E}. The plot of this function in
Fig.\ \ref{fluct:fig}(b) clearly shows that the slope of the function
$G$ is linear up to the distances\footnote{At $R \sim 4 d$ the
deviation of the classical energy from the linear behavior
\eq{Efl:short} is around 15\%.} $R \sim 4 d$ while the density
profile, Fig.\ \ref{fig:density}(d), shows no presence of the string
at $R = 4 d$. This is another evidence in favor of the fact that the
linear term is not caused by the string formation mechanism.

\section{Conclusions}

We studied the static potential between oppositely
charged electric particles in the pointlike magnetic dipole gas. We
found that the short distance potential contains a linear piece due
to overlapping of the dipole clouds gathered near electric sources
because of the induced magnetic field. This effect can not be
explained by a short string formation contrary to the $3D$ compact
electrodynamics~\cite{ChGuPoZa:GG} and $4D$ (dual) Abelian Higgs
model~\cite{GuPoZa98-99}. The coefficient in front of the linear
term is proportional to the density of the dipoles and to the
electric charge of the test particles.

As the distance between the test particles increases the potential
becomes of a Coulomb type. At large distances the dipole effects lead
to the electric charge renormalization.

These results may have interesting applications for physics of gauge
theories both in high and low temperature regimes. One of the
physically interesting candidates of such theories is the electroweak
model in a low temperature phase in which the formation of the
monopole--antimonopole pairs has been observed~\cite{EW}. The dipole
effects may also induce a short distance linear
potential~\cite{BaChVe00} between electric charges in a high
temperature phase of the compact $U(1)$ lattice gauge theory
in three space-time dimensions.

\acknowledgments

The author acknowledges the kind hospitality of the staff of the
Department of Physics and Astronomy of the Vrije University at Amsterdam,
where the work was done. The author is grateful to B.~L.~G.~Bakker,
F.~V.~Gubarev and V.~I.~Shevchenko for useful discussions. This work
was partially supported by the grants RFBR 99-01-01230a and INTAS
96-370.

\newpage

\begin{figure}
\caption{The Wilson loop for the static charge and anti-charge
separated by the distance $R$.}
\label{WLplane}
\end{figure}

\begin{figure}
\caption{(a) The distribution function $F$, eq.\eq{distr}, for the
dipole parameter $r$, and (b)
corresponding (normalized) energy $G(x)$, $x = R \slash d$, of the
charge--anticharge pair, eqs.(\ref{E}).}
\label{fluct:fig}
\end{figure}

\begin{figure}
\caption{The normalized energy density \eq{energy:density} of the
dipole cloud around the electric sources in the plane $(x_1,x_3)$.
The profile and contour plots are shown for various charge
separations:  (a) $R=0.5 d$, (b) $R=1.0 d$, (c) $R=2.0 d$, (d) $R=4.0
d$. The distances are shown in units of $d$.}
\label{fig:density}
\end{figure}

\newpage

\begin{figure}
\centerline{\epsfig{file=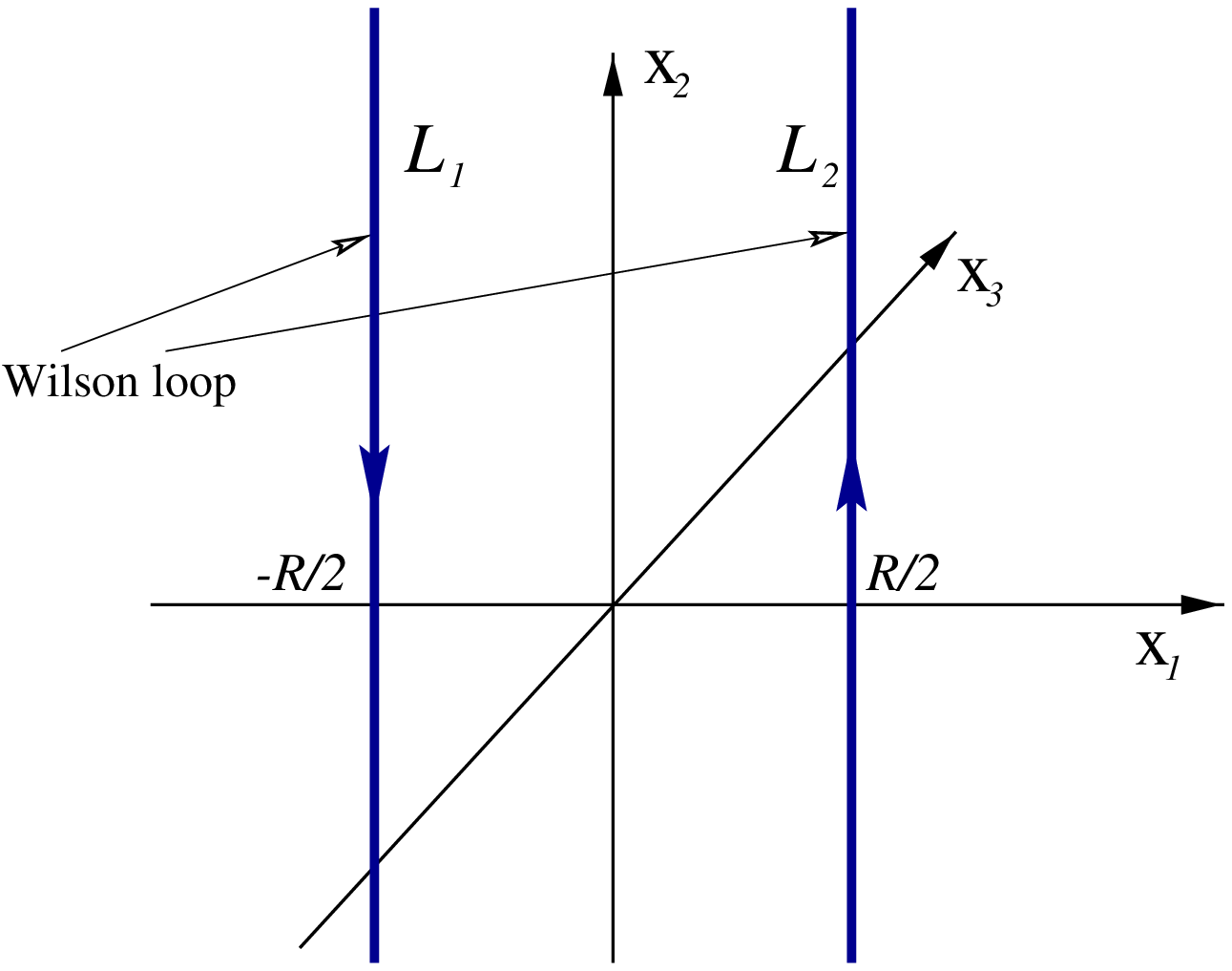,width=10.0cm,height=10.4cm,angle=-90}}
\end{figure}
\centerline{FIG. \ref{WLplane}}

\newpage

\begin{figure}
\centerline{\epsfig{file=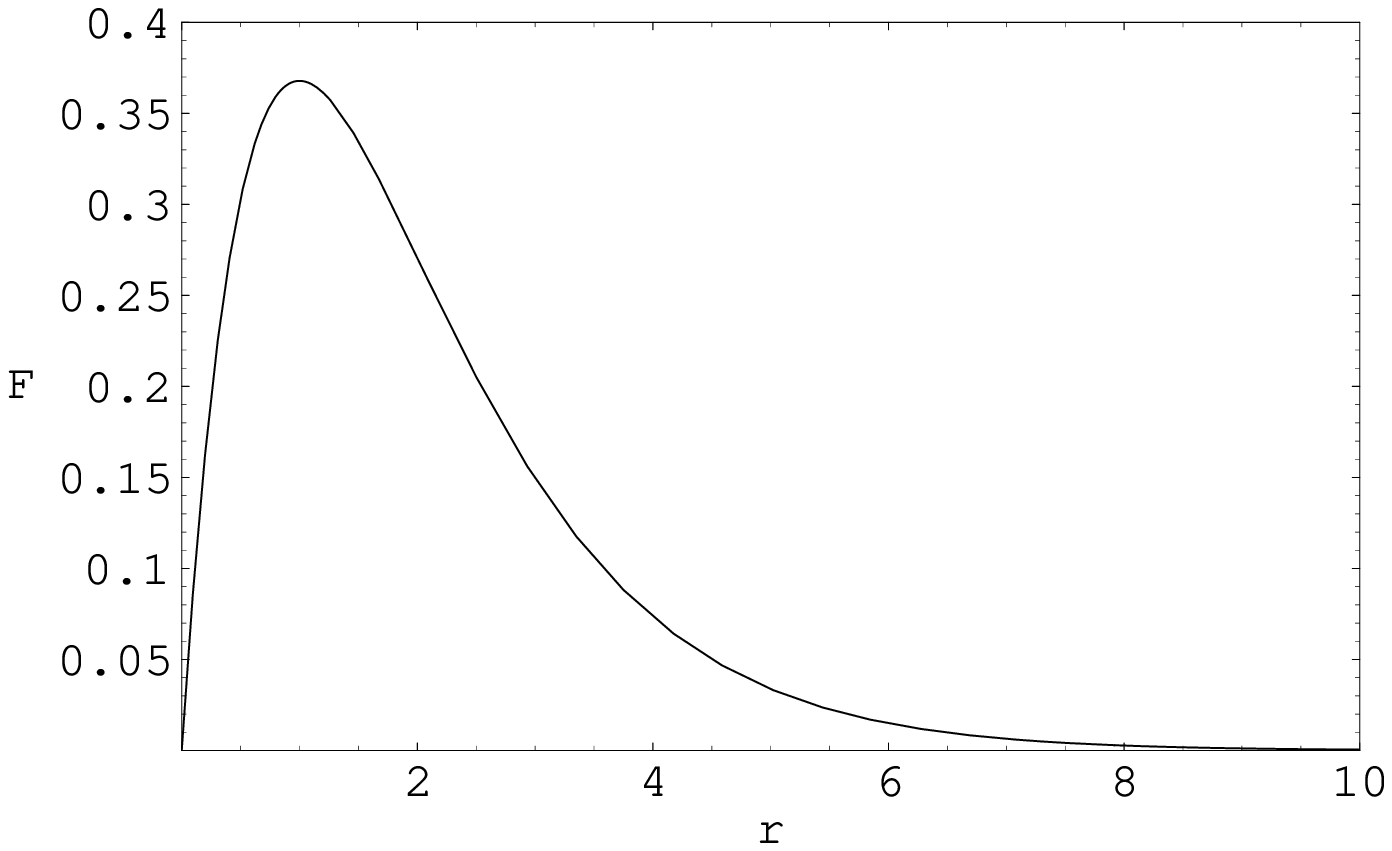,width=14.4cm,height=12.0cm}}
\end{figure}
\centerline{FIG. \ref{fluct:fig}(a)}

\newpage

\begin{figure}
\centerline{\epsfig{file=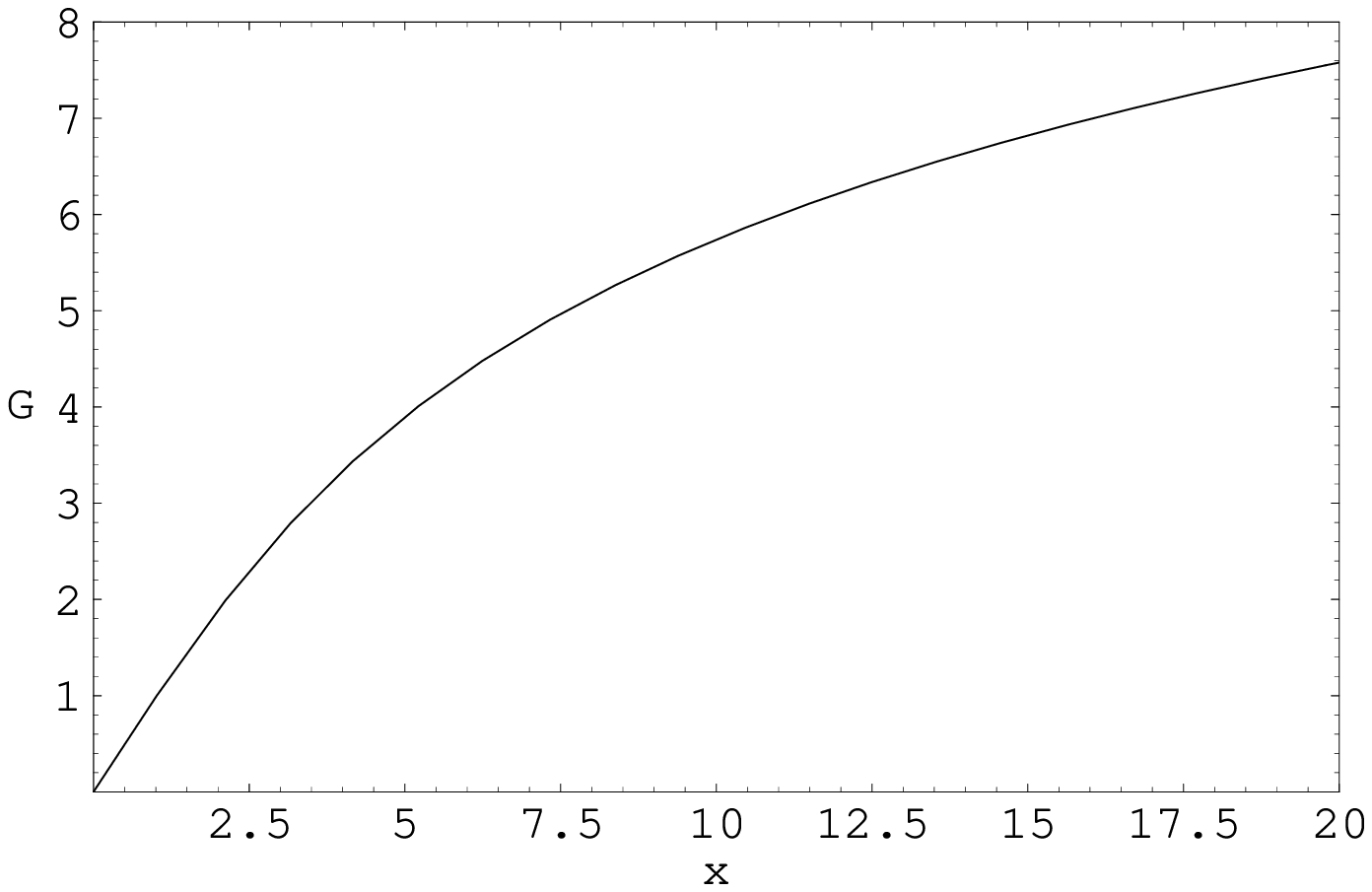,width=14.4cm,height=12.0cm}}
\end{figure}
\centerline{FIG. \ref{fluct:fig}(b)}

\newpage

\begin{figure}
 \begin{center}
  \epsfig{file=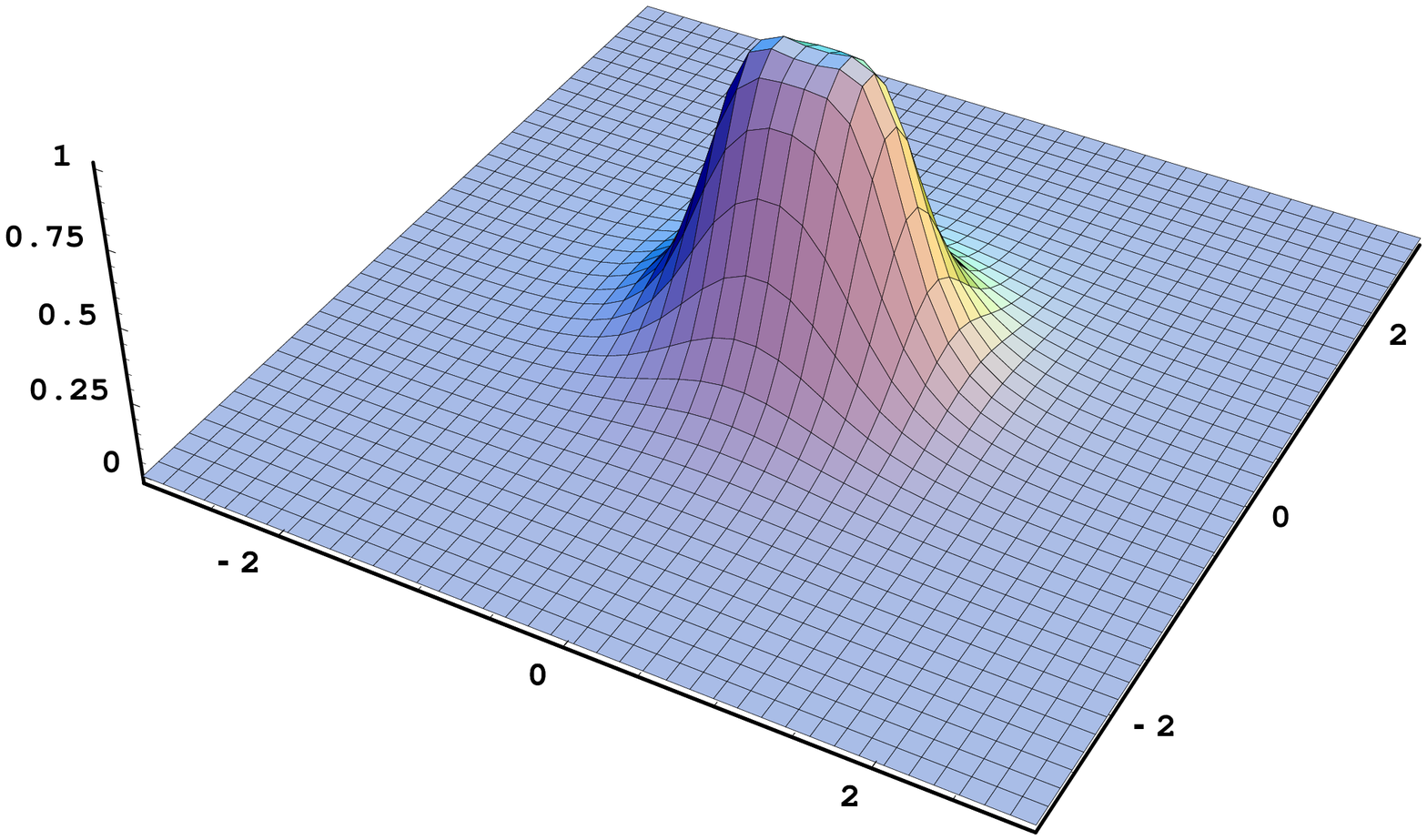,width=14.4cm,height=12.0cm} \\
\vskip 20mm
  \epsfig{file=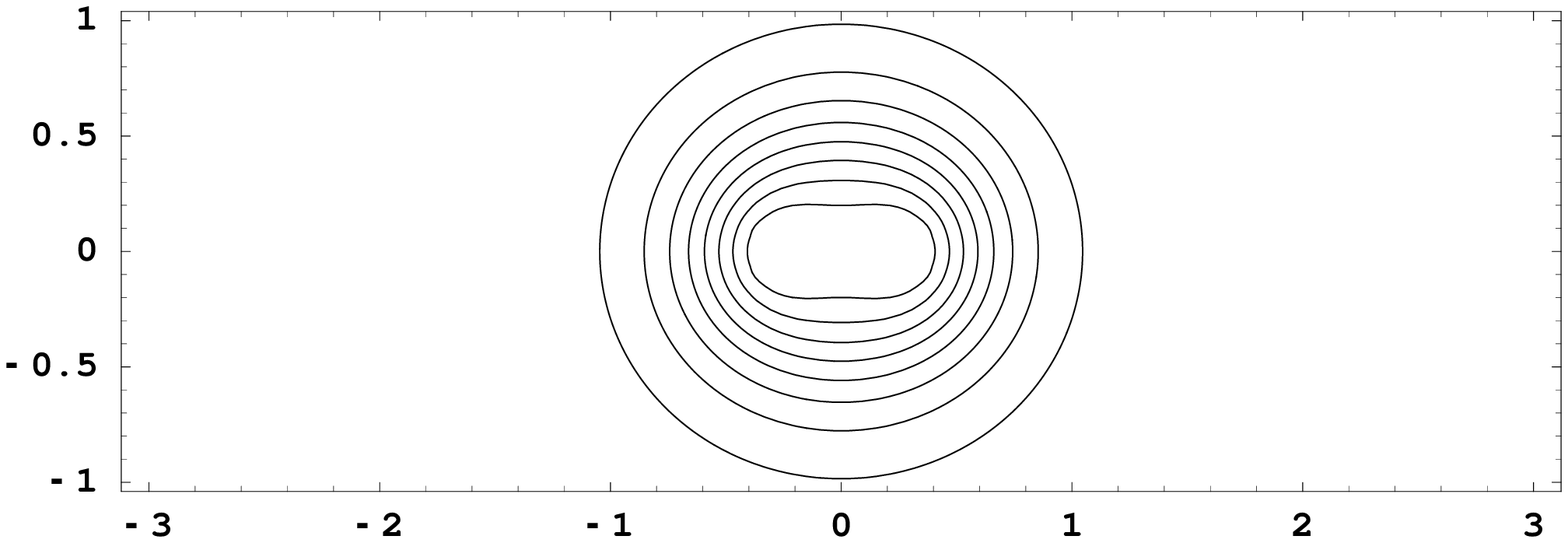, width=14.0cm,height=6.0cm}
 \end{center}
\end{figure}
\centerline{FIG. \ref{fig:density}(a)}

\newpage

\begin{figure}
 \begin{center}
  \epsfig{file=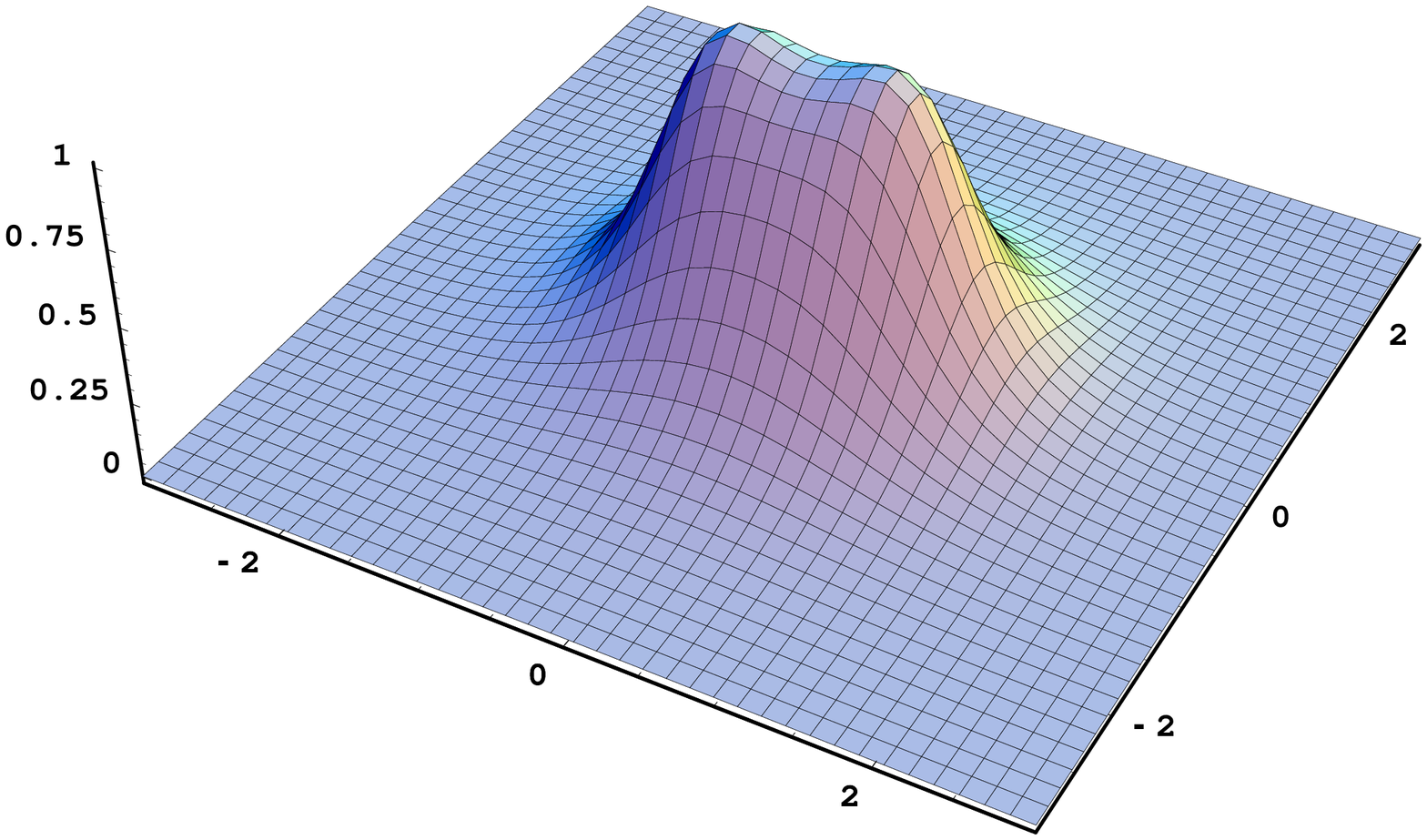,width=14.4cm,height=12.0cm} \\
\vskip 20mm
  \epsfig{file=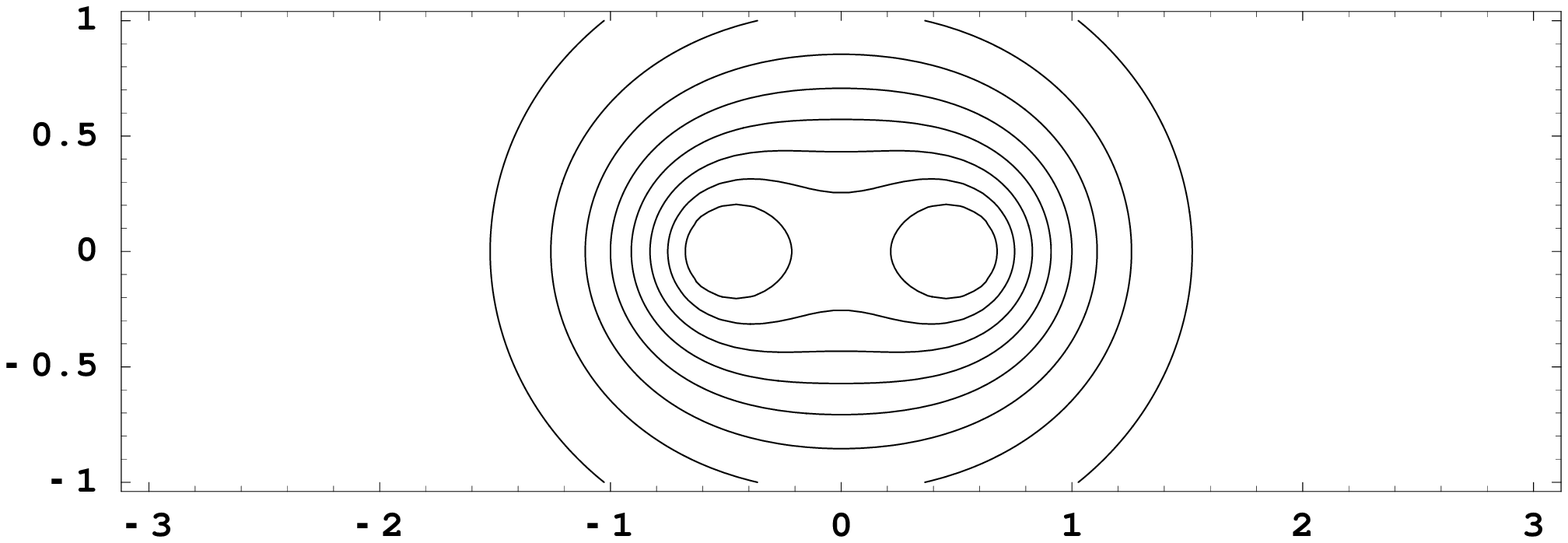, width=14.0cm,height=6.0cm}
 \end{center}
\end{figure}
\centerline{FIG. \ref{fig:density}(b)}

\newpage

\begin{figure}
 \begin{center}
  \epsfig{file=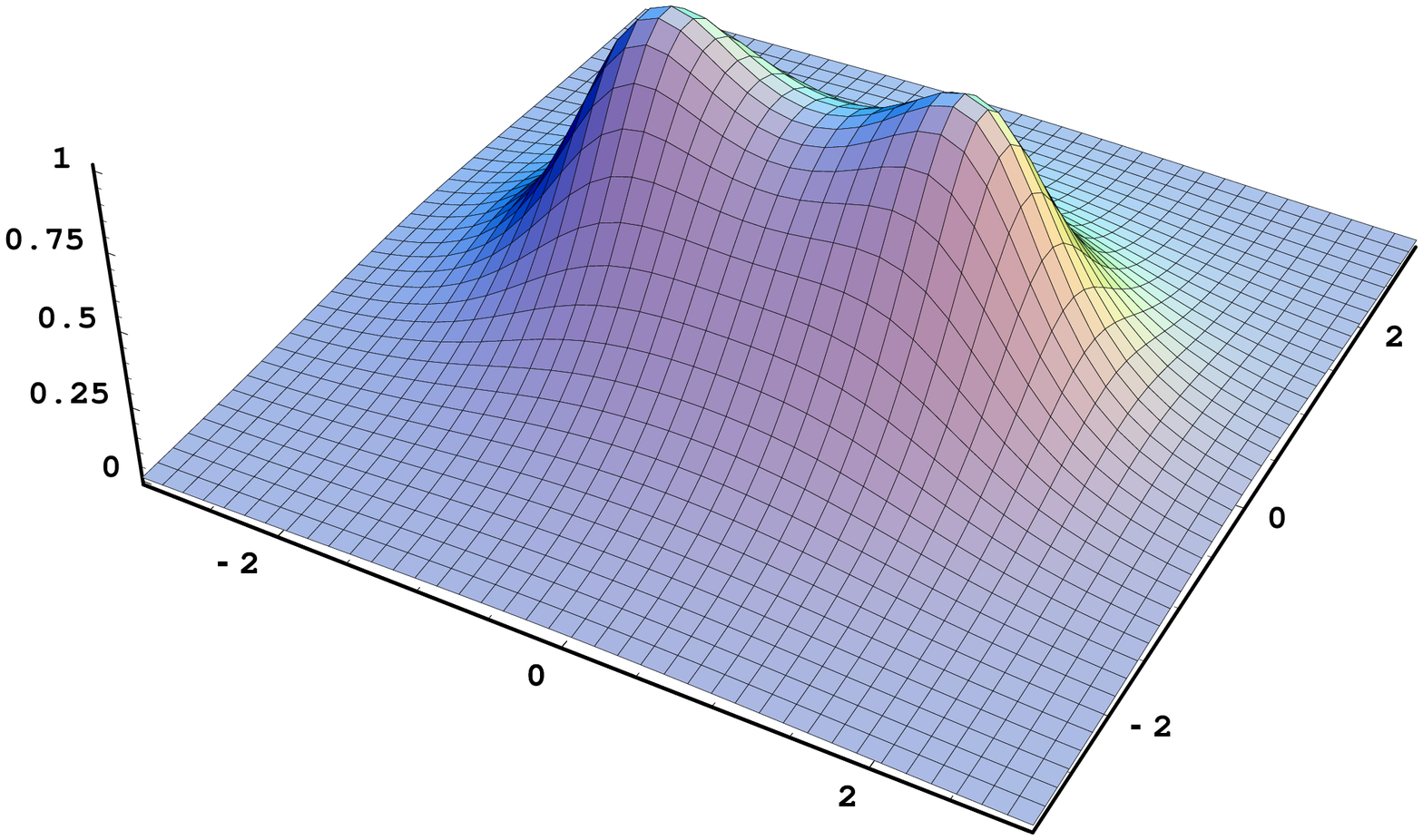,width=14.4cm,height=12.0cm} \\
\vskip 20mm
  \epsfig{file=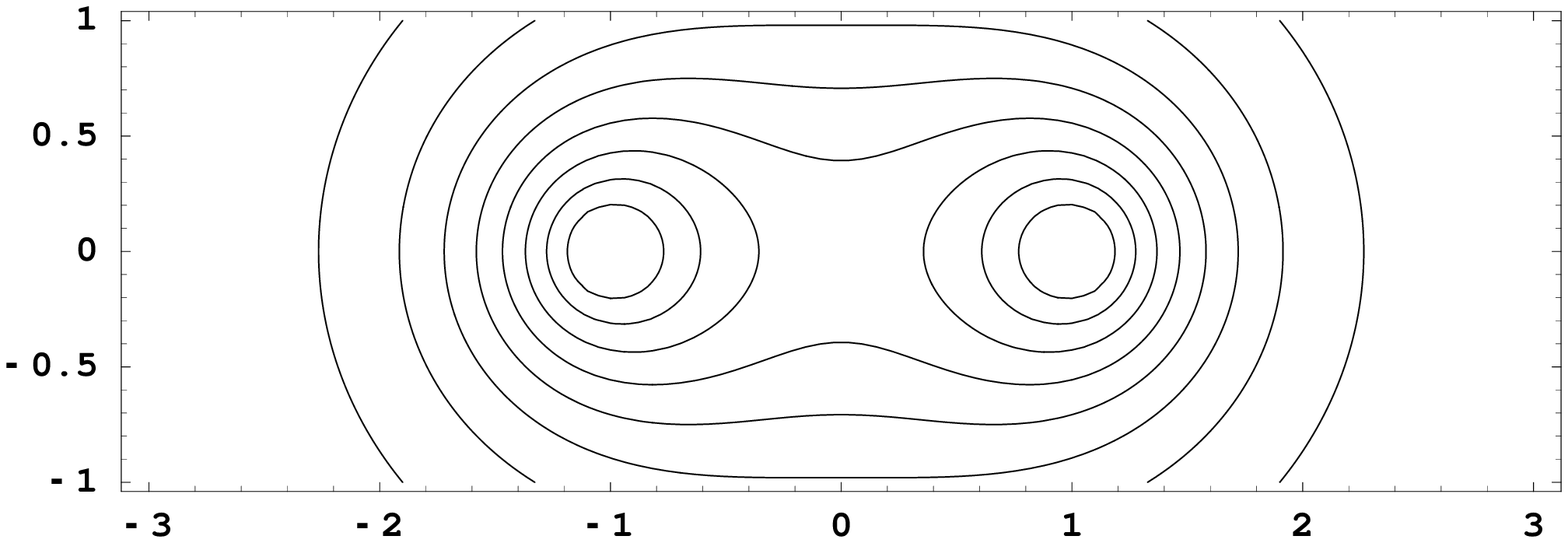, width=14.0cm,height=6.0cm}
 \end{center}
\end{figure}
\centerline{FIG. \ref{fig:density}(c)}

\newpage

\begin{figure}
 \begin{center}
  \epsfig{file=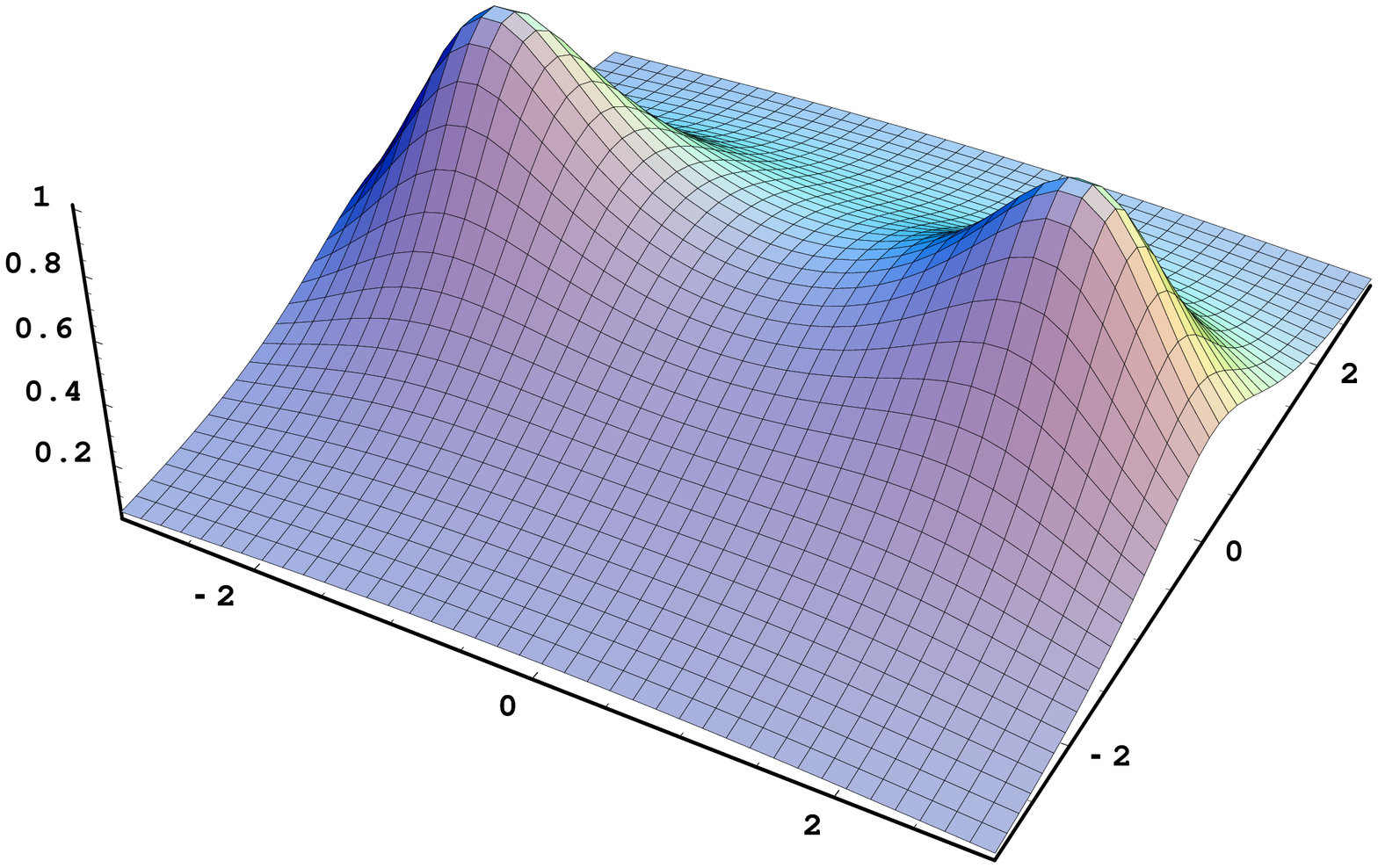,width=14.4cm,height=12.0cm} \\
\vskip 20mm
  \epsfig{file=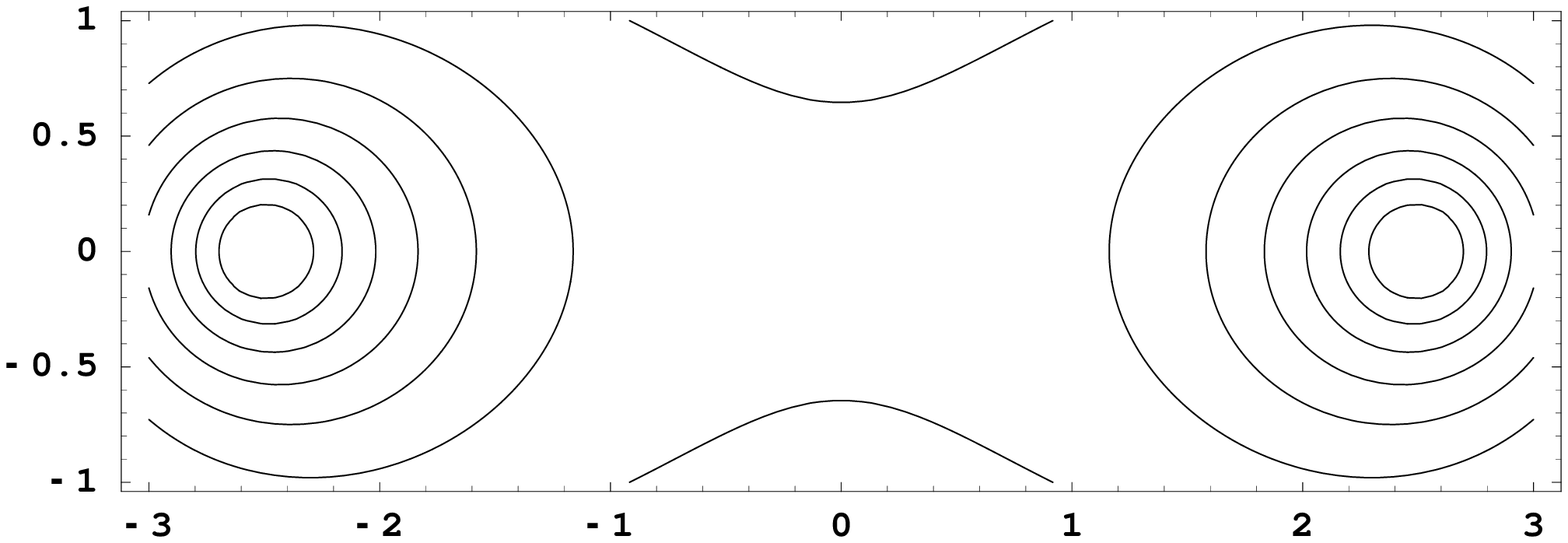, width=14.0cm,height=6.0cm}
 \end{center}
\end{figure}
\centerline{FIG. \ref{fig:density}(d)}


\begin{thebibliography}{99}

\bibitem{tHPo} G.~'t~Hooft, Nucl.~Phys. B79, 276 (1974).\\
A.~M.~Polyakov,  JETP Lett. 20, 194 (1974).

\bibitem{Polyakov} A.~M.~Polyakov, Nucl.~Phys. B120, 429 (1977).

\bibitem{ChGuPoZa:GG} M.~N.~Chernodub, F.~V.~Gubarev, M.~I.~Polikarpov
and V.~I.~Zakharov, Phys.~Lett. B475, 303 (2000).

\bibitem{VIZandCo}
R.~Akhoury, V.I.~Zakharov, Phys. Lett. B438, 165 (1998).\\
K.~G.~Chetyrkin, S.~Narison, V.~I.~Zakharov, Nucl. Phys. B550, 353
(1999).

\bibitem{GuPoZa98-99}
F.V.~Gubarev, M.I.~Polikarpov, V.I.~Zakharov,  Phys. Lett. B438,
 147 (1998); Mod. Phys. Lett. A14 (1999) 2039; hep-ph/9908292.


\bibitem{B} V.~L.~Berezinsky,  Sov.~Phys.~JETP 32, 493 (1971)

\bibitem{KT} J.~M.~Kosterlitz and D.~J.~Thouless, J.~Phys. C6, 1181
(1973).

\bibitem{FrSp80} J.~Fr\"ohlich and T.~Spencer, J.~Stat.~Phys.
24, 617 (1981).

\bibitem{AgasianZarembo} N.~O.~Agasian and K.~Zarembo,
Phys.~Rev. D57, 2475 (1998).

\bibitem{no-screening} J.~Glimm and A.~Jaffe, Comm.~Math.~Phys. 56,
195 (1977).

\bibitem{Antonov} D.~Antonov, Int.~J.~Mod.~Phys. A14, 4347 (1999).

\bibitem{Bogomolny} E.~B.~Bogomol'ny, Sov.~J.~Nucl.~Phys. 24, 449
(1976).

\bibitem{ANO} A.~A.~Abrikosov, Sov.~Phys.~JETP 5, 1174 (1974).\\
H.~B.~Nielsen and P.~Olesen, Nucl.~Phys. B61, 45 (1973).

\bibitem{EW}
M.~N.~Chernodub, F.~V.~Gubarev and E.--M.~Ilgenfritz,
Phys.~Lett. B424, 106 (1998).\\
M.~N.~Chernodub, F.~V.~Gubarev, E.--M.~Ilgenfritz and A.~Schiller,
Phys.~Lett. B443, 244 (1998); B434, 83  (1998).

\bibitem{BaChVe00} B.~L.~G.~Bakker, M.~N.~Chernodub and A.~I.~Veselov
(in preparation).

\end{thebibliography}
\end{document}